\documentclass{aa}

\usepackage{hyperref}
\usepackage{placeins}

\usepackage{cleveref}

\usepackage{graphicx}

\usepackage{booktabs}
\usepackage{multirow}
\usepackage{caption}
\usepackage{etoolbox}
\usepackage{latexsym}
\usepackage{longtable}
\usepackage{tabularx}
\usepackage[table]{xcolor}
\definecolor{MidnightBlue}{RGB}{25, 25, 112}
\definecolor{MidnightBlueLight}{RGB}{239,239,251}

\usepackage{subcaption} 

\usepackage{txfonts}

\begin{document} 

\titlerunning{CCSN detections from ET}
\title{Core-collapse supernova detections from Einstein Telescope within the Milky Way}

\author{
I. F. Giudice \thanks{\email{ines.giudice@inaf.it,alessio.desantis@gssi.it}}
\inst{1} 
\and
A. L. De Santis*
\inst{2, 4}
\and
M. T. Botticella
\inst{1}
\and
M. Branchesi
\inst{2, 4}
\and
G. Pastorelli
\inst{4}
\and
L. Girardi
\inst{4}
\and
L. Izzo
\inst{1,3}
\and 
E. Cappellaro
\inst{4}
\and
M. Della Valle
\inst{4}
      } 

\institute{INAF, Osservatorio Astronomico di Capodimonte, Salita Moiariello 16, I-80131, Naples, Italy  
\and Gran Sasso Science Institute (GSSI), I-67100 L’Aquila, Italy
\and DARK, Niels Bohr Institute, University of Copenhagen, Jagtvej 128, 2200 Copenhagen, Denmark
\and INAF, Osservatorio Astronomico di Padova, vicolo dell'Osservatorio 5, 35122 Padova, Italy
\and INFN, Laboratori Nazionali del Gran Sasso, I-67100 Assergi, Italy
       }

  \abstract

   {
   Core-collapse supernovae are promising sources of gravitational waves, which provide a unique probe of the physics driving their explosion mechanism. Third-generation detectors, such as the Einstein Telescope, will dramatically improve the prospects for detecting these signals.
}
   {This study assesses the capability of the Einstein Telescope, alone and in synergy with next-generation detectors such as Cosmic Explorer, to detect gravitational waves from core-collapse supernovae. We estimate the detection horizons and expected event rates for sources in the Milky Way and nearby satellite galaxies.}
   {We employed the \texttt{GWFish} simulation framework, customized to include core-collapse supernovae waveform catalogs from state-of-the-art 3D  simulations and stellar population data generated with \texttt{TRILEGAL}. This approach allows us to model gravitational waves detectability as a function of progenitor mass, source position and detector network configuration.
   }
  {Our analysis shows that the Einstein Telescope can detect gravitational waves from proto-neutron star driven core-collapse supernovae up to distances ranging from $\sim$20 to more than 100 kpc with a 90$\%$ confidence level, depending on the progenitor mass and waveform, while a combination of this detector in a network can extend the reach up to $\sim$170 kpc in the most favorable cases. For a representative 15 M$_{\odot}$ progenitor, ET (in its 2L configuration) achieves a detection horizon of $\sim$100 kpc, ensuring essentially complete coverage of the Milky Way and partial coverage of the Magellanic Clouds. } 
   {}

   \keywords{gravitational waves -- stars: supernovae: general --
          Galaxy: general -- methods: numerical}
   
   \maketitle

\section{Introduction}   
Core-Collapse Supernovae (CCSNe) mark the final stage of massive stars ($\geq$8 M$_{\odot}$) and are among the most energetic explosions in the universe. These cataclysmic explosions disrupt stellar envelopes and result in the formation of compact objects, either neutron stars or black holes \citep{woosley2002,Woosley1995}. These events not only contribute to the synthesis of heavy elements via explosive nucleosynthesis and their injection into the interstellar medium, enriching future generations of stars and planets, but also serve as key probes of fundamental physics under extreme conditions. Moreover, the CCSN rate (CCSNR) provides valuable insights into star formation rates (SFRs) and the chemical enrichment history of galaxies across cosmic time.\\
Our understanding of CCSNe has been primarily shaped by observations across the electromagnetic spectrum, which have yielded crucial information about progenitor stars, explosion energetics and elements nucleosynthesis. However, these observations offer limited insight into the central engine that drives the collapse and explosion, as the core is largely opaque to photons. To probe the internal mechanism, alternative messengers are required such as neutrinos and gravitational waves (GWs). 
The weak interaction of neutrinos allows them to escape from the stellar core within seconds of the collapse, providing an early perspective on core thermodynamics, lepton number evolution and phases such as the bounce and shock revival \citep{janka2017neutrino}.  The observational evidence for this channel was established through the detection of a small number of neutrinos from SN 1987A in the Large Magellanic Cloud \citep{Hirata1987,bionta1991}, which also provided the first direct empirical confirmation of the core-collapse mechanism. 

Similarly, GWs offer a direct probe of the collapse dynamics, multidimensional instabilities, the properties of the forming proto-neutron star (PNS), and potentially the nuclear equation of state \citep{Mezzacappa2024,Richers2017,Jakobus2023}. In initially non-rotating or slowly rotating progenitors, the broadband emission is generated by prompt post-shock convection, neutrino-driven convection, turbulence, the Standing Accretion Shock Instability (SASI), anisotropic accretion flows, and the oscillation modes that these flows excite in the PNS \citep{Andresen:2017,powell2019gw,radice2019characterizing, Mezzacappa2024,vartanyan2023gravitational}. These features can in principle be used to constrain the explosion mechanism and the dynamics of the central engine \citep{Powell2016ExplosionMechanism,Roma2019,Torres2019}. The CCSN GW spectrum covers a large band from roughly mHz to kHz. Slowly varying matter and neutrino asymmetries generate a so called memory component predominantly below tens of Hz and, for sufficiently long signals, mHz frequencies. Stochastic matter motions and PNS oscillations produce emission from tens of Hz to several kHz, with much of the GW energy in neutrino-driven simulations lying between approximately 100 and 2000\,Hz \citep{VartanyanBurrows2020,Richardson2022,vartanyan2023gravitational,Choi2024}. Detection is challenging because of the weak strain, the stochastic nature of many of the signal components and as parts of the signal lie outside the current bands of ground-based interferometers. The advent of third-generation observatories such as the Einstein Telescope (ET) is expected to overcome this limitation. ET is the upcoming European ground-based GW observatory designed to achieve an order of magnitude improvement in strain sensitivity with respect to the current generation and to extend the observational band down to a few hertz \citep{punturo2010einstein,abac2025science}.  This sensitivity improvement will enable the detection of tens of thousands of compact binaries out to high redshift, as well as binary neutron star systems during their inspiral phase before merger. It will also enhance the detectability of weak transient sources, such as CCSNe. Two observatory design are currently under investigation: a triangular configuration consisting of three nested 10\,km interferometers ($\Delta$) and a dual-L (2L) configuration composed of two 15\,km interferometers \citep{branchesi2023science}. Both designs provide an increase in astrophysical reach with respect to current facilities, although they differ in their sensitivity, localization capabilities and response to different classes of GW sources. Beyond increasing the number of detectable GW sources, ET is expected to address several major questions in astrophysics and fundamental physics. Its unprecedented sensitivity will enable the observation of a wider population of compact-object mergers, providing new insights into the formation and evolution of black holes and neutron stars. This increase in binary coalescence detection  will also serve as cosmological probes through the use of GW standard sirens, allowing independent measurements of the expansion history of the Universe. In addition, ET will offer unique opportunities to test General Relativity in the strong field regime, constrain the equation of state of ultra-dense matter, investigate the stochastic GW background and search for signals from currently unexplored classes of astrophysical transients \citep{abac2025science,branchesi2023science}. Among the transient sources that will particularly benefit from this increased sensitivity are Galactic CCSNe. Events occurring within the Milky Way (MW) offer the rare opportunity to detect all astrophysical messengers (electromagnetic radiation, neutrinos, GWs and possibly cosmic rays) enabling a comprehensive multimessenger study. 
Within the MW, the negligible interactions of GWs and neutrinos are essential as massive star progenitors are concentrated in the dusty Galactic plane. THerefore a substantial fraction of Galactic CCSNe may be highly extinguished or entirely missed by optical observations \citep{adams2013observing,murphey2021witnessing}. GW and neutrino observations can detect a collapse independently of optical extinction, provide an early trigger, and guide infrared, radio, X-ray and gamma-ray follow-up \citep{Nakamura2016Multimessenger,Al_Kharusi_2021,powell2025}. 

We investigated the prospects for detecting CCSNe within the MW with the ET by combining state-of-the-art PNS-driven CCSN waveform models with a realistic Galactic progenitor population generated using the \texttt{TRILEGAL} stellar population synthesis code  \citep{girardi2005star,girardi2012trilegal}. The analysis was restricted to progenitors with masses in the range of $9-25\,M_{\odot}$, corresponding to the coverage of the CCSN waveform catalog adopted in this work. Restricting the progenitor sample to this interval ensures consistency between the simulated Galactic population and the available GW signal models. We constructed a realistic Galactic progenitor population and derived a self-consistent estimate of the Galactic CCSNR. We then quantified detection horizons and observational coverage achievable with different ET configurations, both operating alone and in combination with Cosmic Explorer (CE). To perform this analysis, we extended the \texttt{GWFish} framework \citep{dupletsa2023gwfish} by implementing CCSN waveform models and realistic Galactic progenitor populations. Taking into account the increased CCSN detection horizon provided by networks of GW observatories, we expanded our analysis to nearby galaxies, including the Large and Small Magellanic Clouds.
We further investigated the impact of Galactic dust extinction on multimessenger observations, assessing the fraction of CCSNe that may remain hidden from optical surveys while still being detectable through GW. This framework enables realistic forecasts that are particularly relevant for the development of future multimessenger observing strategies. In this context,  next generation GW and  neutrino detectors can serve as triggers for rapid multi-wavelength follow-up with optical, radio, X-ray and gamma-ray facilities. 

This paper is organized as follows. In Section \ref{rates}, we review current estimates of the CCSNR. Then in Section\,\ref{sec:progenitors}, we describe the Galactic progenitor population from the \texttt{TRILEGAL} simulation. We then present our estimate derived from the simulated progenitor distribution in Section \ref{sec:ccsnrsfr}. Sections \ref{sec:gw} and \ref{sec:gwfish} describe the gravitational waveform models and the simulation framework used to assess the detection capabilities of ET. We present our main findings, including detection horizons, in Section \ref{sec:results}. Lastly we discuss the implications of these results in the context of multi-messenger astronomy in Section \ref{sec:discussion}, and provide concluding remarks in Section \ref{sec:conclusion}.

\section{CCSNR within the Milky Way}\label{rates}
The CCSNR within the MW is a crucial parameter for optimizing observational strategies of current and future GW and neutrino detectors. No Galactic SN has been optically observed since 1604, primarily due to dust extinction in the Galactic plane. Nevertheless, studies of SN remnants suggest that recent core-collapse events may have occurred, with G1.9+0.3 being the youngest known remnant, dated to approximately 110–180 years ago \citep{reynolds2018evolution, luken2020radio}. Historical records \citep{Stephenson2002} provide direct evidence for only one confirmed Galactic CCSN, namely SN 1054, which was likely a Type IIn-P event \citep{smith2013crab}. Although SN 1181 was historically associated with the core-collapse pulsar wind nebula 3C 58 \citep{kothes2013distance}, its classification remained a subject of intense debate for decades and recent multi-wavelength investigations have  linked SN 1181 to the nebula Pa 30 and its central star, reclassifying it as a sub-luminous, thermonuclear Type Iax supernova \citep{Schaefer2023}.
Multiple approaches have been employed to estimate the Galactic CCSNR, each based on different observational data and modeling assumptions. 
One approach combines historical SNe with models of Galactic extinction and stellar distribution to infer the likelihood of past detections. Using this method, \cite{adams2013observing} performed Monte Carlo simulations adopting a Salpeter \citep{Salpeter} initial mass function (IMF) and progenitor masses between 8–100 $M_{\odot}$, deriving a CCSNR of $3.2^{+7.3}_{-2.6}$ (100 yr)$^{-1}$. Similar analyses by \cite{murphey2021witnessing}, modeling the visibility of naked-eye SNe and accounting for Galactic dust, suggest that about 33$\%$ of CCSNe would have been observable, yielding a rate of $1.4^{+1.6}_{-0.9}$ (100 yr)$^{-1}$. Another strategy extrapolates the CCSNR from external galaxies with similar morphology and star formation properties. Early estimates by \cite{cappellaro1993rate} and \cite{van1994rediscussion}, assuming the MW resembles an Sb–Sbc galaxy, yielded rates of $1.4 \pm 1.0$ and $3.0 \pm 1.0$ (100 yr)$^{-1}$, respectively. More recent analyses using larger galaxy samples, such as \cite{li2011nearby}, report $2.30 \pm 0.48$ (100 yr)$^{-1}$. While sensitive to the assumed Galactic type and luminosity, these extragalactic estimates are broadly consistent. The CCSNR can also be inferred from the birthrate of massive stars. \cite{reed2005new} analyzed a census of O3–B2 dwarfs within 1.5 kpc of the Sun, adopting a “disk+hole” density model, and estimated a CCSNR of 1–2 (100 yr)$^{-1}$. Following this approach, \cite{quintana2025census} conducted a census of OB stars within 1 kpc, using the same plane density function ($\rho_{\rm plane}$) as \cite{reed2005new}. Their extrapolated CCSNR is $0.5 \pm 0.1$ (100 yr)$^{-1}$, roughly half of the previous estimate. This discrepancy is due to the smaller sample size and to differences in the stellar mass range and evolutionary models used, for instance \cite{reed2005new} included O3–B2 stars and non-rotating models, shortening stellar lifetimes.
The birthrate of neutron stars provides a further independent constraint. Syntheses of neutron star populations, including pulsars, magnetars, XDINSs and RRATs, combined with the NE2001 Galactic electron density model, yield a rate of $\sim$7.2 events (100 yr)$^{-1}$ \citep{rozwadowska2021rate}. Then, gamma-ray emission from radioactive isotopes, such as $^{26}$Al mapped by INTEGRAL \citep{diehl2006radioactive}, traces recent nucleosynthetic activity and thus CCSNe. Assuming theoretical yields and a Scalo IMF, the inferred CCSNR is $1.9 \pm 1.1$ (100 yr)$^{-1}$. By combining these independent approaches (stellar population studies, NS birthrates, extragalactic scaling and radioactive isotope constraints) \cite{rozwadowska2021rate} obtained a Galactic CCSNR of $1.63 \pm 0.46$ (100 yr)$^{-1}$. This value infers multiple proxies, hence balancing the uncertainties and biases inherent in each.

\section{Progenitor simulation}\label{sec:progenitors}

To investigate the CCSN detectability across our Galaxy, we analyzed an all-sky distribution of massive stars simulated using the TRIdimensional modeL of thE GALaxy (\texttt{TRILEGAL}) code \citep{girardi2005star, girardi2012trilegal}. {\tt TRILEGAL} is a population synthesis tool that generates stellar catalogs based on a comprehensive model of the Galaxy’s structure (incorporating the thin disk, thick disk, bulge, and halo), with options to simulate external systems such as the Magellanic Clouds. The code combines realistic assumptions for the star formation history, IMF, age-metallicity relations and stellar evolution to produce a spatially resolved stellar population. For the all-sky Milky Way simulation, we compute a dedicated catalog using the updated Galactic disc SFH derived by \citep{mazzi2024dissecting} and to ensure completeness in the sample, especially for faint progenitors, we set the bolometric magnitude cut to $M_{bol} < 30$ mag. This threshold ensures the inclusion of progenitor stars that might be undetectable in optical surveys, yet remain relevant for the GW source population. In addition, no extinction was considered in the calculated apparent magnitudes. For both the Milky Way and the Magellanic Clouds, we only consider stars from single stellar evolution models, namely PARSEC v1.2S (\citep{bressan2012parsec,tang,chen2015} and COLIBRI (\citep{marigo2013evolution,rosenfield2016evolution,pastorelli2019,pastorelli2020}, and we assumed a Kroupa IMF \citep{kroupa2001variation}. In our analysis we selected stars with initial masses between 9 and 25 M$_{\odot}$. The simulated catalogs provide the spatial distribution of these massive stars, along with other parameters such as masses, ages and distances. 

\section{CCSNR estimate using the SFR}
\label{sec:ccsnrsfr}

An independent estimate of the Galactic CCSNR can be obtained by exploiting the relation between the SFR and the CCSNR, since the short lifetimes of massive stars imply that the CCSNR closely traces the recent SFR. Under the assumption of a constant IMF and negligible delay time between star formation and explosion, the CCSNR can be approximated as

\begin{equation}
\mathrm{CCSNR}(t) = K_{\mathrm{CC}}\,\psi(t),
\end{equation}

where $\psi(t)$ is the SFR and $K_{\mathrm{CC}}$ is a scaling factor. $K_{\mathrm{CC}}$ depends on the adopted IMF and on the progenitor mass range capable of producing CCSNe and is defined as:
\begin{equation}
	K_{\mathrm{CC}} = \frac{\int_{m_{l}^{CC}}^{m_{u}^{CC}} \phi(m,\tau)\,dm}{\int_{m_{l}}^{m_{u}} m\,\phi(m,\tau)\,dm}
	\label{eq:ccsnr}
\end{equation}
where $\phi(m,\tau)$ is the IMF, $m_{l}-m_{u}$ is the mass range for the IMF and $m_{l}^{CC}-m_{u}^{CC}$ is the suitable mass range for CCSN. 

We estimated the CCSNR by adopting the SFR implemented in \texttt{TRILEGAL} \citep{mazzi2024dissecting}, which provides a spatially resolved description of the Galactic stellar populations and their formation history. For the IMF, we adopt the Kroupa IMF \citep{kroupa2001variation}, which is widely used in Galactic stellar population studies and is also implemented in \texttt{TRILEGAL}.  By selecting the following mass ranges $m_{l}-m_{u}=0.1-100$\,M$_{\odot}$ and $m_{l}^{CC}-m_{u}^{CC}=9-25$\,M$_{\odot}$ , we obtained a $K_{CC}=0.007$ $M_{\odot}^{-1}$ which leads to a CCSNR of 1.9 (100 yr)$^{-1}$. 
Using this framework, the CCSNR is directly derived from the SFR provided by the \texttt{TRILEGAL} simulation. This approach ensures consistency between the CCSNR estimate and the spatial distribution of massive stars used in our detectability analysis. This result is consistent with selected Galactic-rate estimates discussed in Section \ref{rates}, especially with \cite{rozwadowska2021rate}. The full set of estimates that can be seen in Table~\ref{tab:ccsn_rates} spans $0.5-7.2\,(100\,{\rm yr})^{-1}$, reflecting different tracers and systematic assumptions. This reinforces the reliability of our \texttt{TRILEGAL} based approach as a representative framework for the Galactic CCSN population.

\begin{table}[ht]
\centering
\caption{Summary of Galactic CCSNR estimates.}
\label{tab:ccsn_rates}
\begin{tabular}{lc}
\toprule
Method / Reference & CCSNR (100 yr)$^{-1}$ \\
\midrule

Historical SNe \& extinction & \\
\quad\citet{adams2013observing} & $3.2^{+7.3}_{-2.6}$ \\
\quad\citet{murphey2021witnessing} & $1.4^{+1.6}_{-0.9}$ \\
\midrule

Extragalactic scaling & \\
\quad\citet{cappellaro1993rate} & $1.4 \pm 1.0$ \\
\quad\citet{van1994rediscussion} & $3.0 \pm 1.0$ \\
\quad\citet{li2011nearby} & $2.30 \pm 0.48$ \\
\midrule

OB-star census & \\
\quad\citet{reed2005new} & $1$--$2$ \\
\quad\citet{quintana2025census} & $0.5 \pm 0.1$ \\
\midrule

Neutron-star birthrate & \\
\quad\citet{rozwadowska2021rate} & $\sim7.2$ \\
\midrule

$^{26}$Al $\gamma$-ray emission & \\
\quad\citet{diehl2006radioactive} & $1.9 \pm 1.1$ \\
\midrule

Combined estimate & \\
\quad\citet{rozwadowska2021rate} & $1.63 \pm 0.46$ \\
\midrule
This work & 1.9 \\
\bottomrule
\end{tabular}
\end{table}

\section{Gravitational waves from CCSNe and waveform sample}
\label{sec:gw}

\begin{figure}[!ht]
    \centering
    \begin{subfigure}[t]{\columnwidth}
        \centering
        \includegraphics[width=\columnwidth]{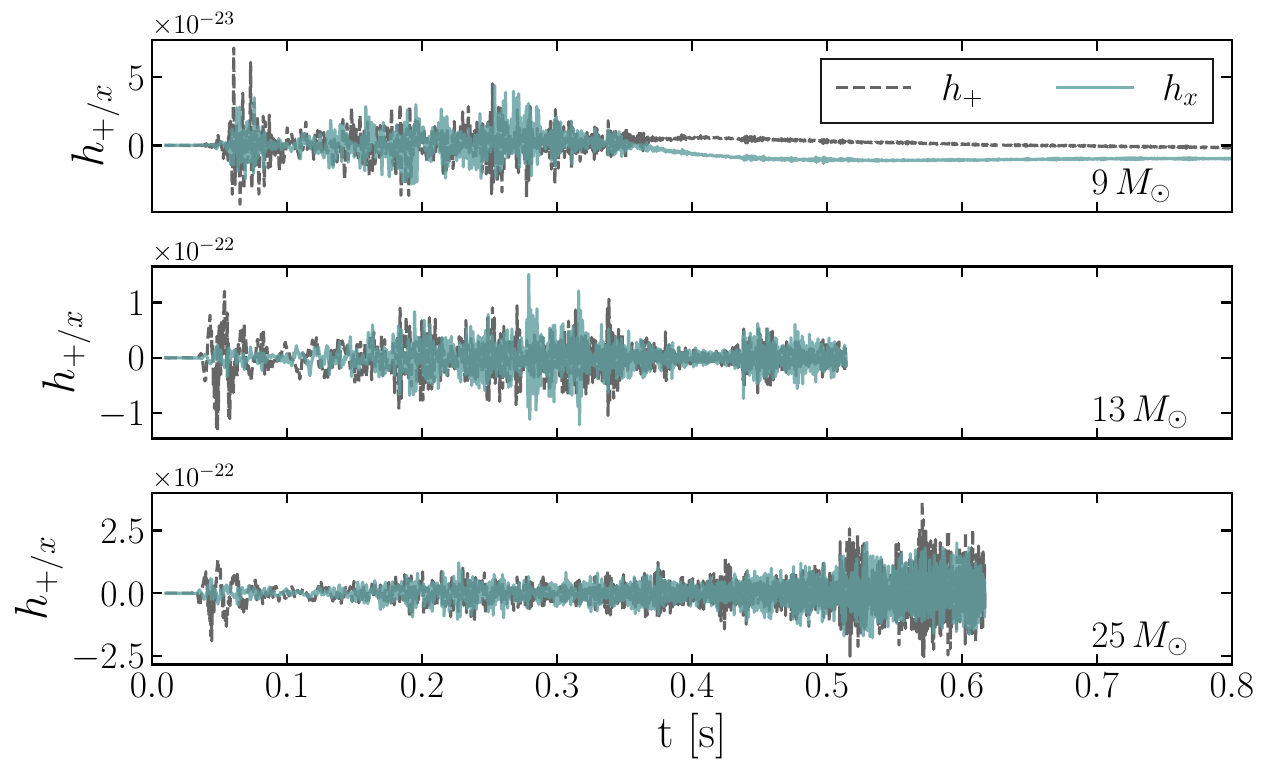}
        \caption{Waveforms simulated from \cite{radice2019characterizing}.}
        \label{fig:strainrad}
    \end{subfigure}
    \vspace{1em}
    \begin{subfigure}[t]{\columnwidth}
        \centering
        \includegraphics[width=\columnwidth]{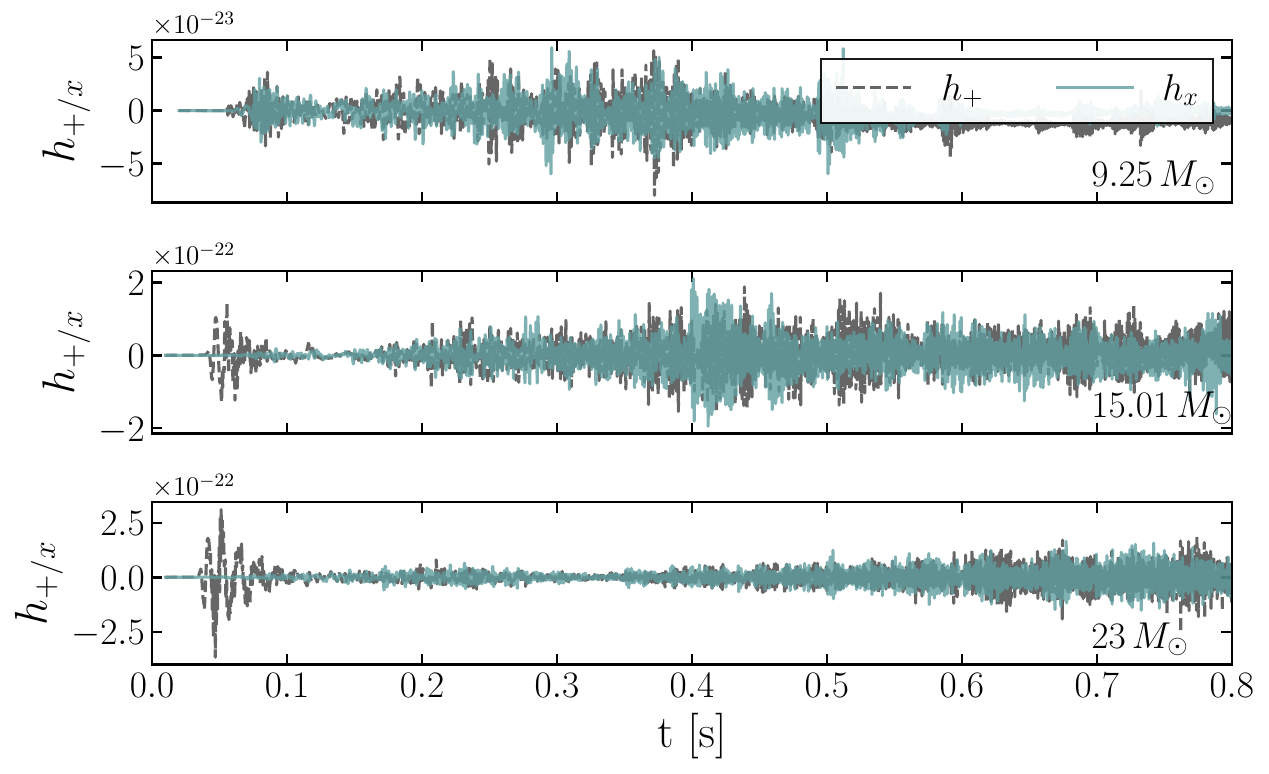}
        \caption{Waveforms simulated from \cite{vartanyan2023gravitational}}
        \label{fig:strainvar}
    \end{subfigure}
    \caption{Representative strains used, scaled to a source distance of 10\,kpc. For each simulation set, we present three progenitors representing the low- and high-mass ends and an intermediate mass of our selected range.}
    \label{fig:strains}
\end{figure}

GWs from CCSNe originate from time-dependent asymmetries in the collapsing stellar core. When the central density of the collapsing star reaches nuclear values the equation of state stiffens. This stiffening causes a rebound of the inner core and launches a shock. In a non-rotating or slowly rotating progenitor this process remains nearly spherical and produces only a weak GW signal. Instead, for a rapidly rotating collapse the centrifugal forces lead to a pronounced short-duration core-bounce burst. Its characteristic frequency is generally several hundred hertz to approximately one kilohertz and depends on both the rotation profile and the nuclear equation of state \citep{Dimmelmeier2008,Richers2017}. For the initially non-rotating FORNAX models adopted here the emission visible during the first $\sim0.1\,\rm s$ in Fig.~\ref{fig:strains}
is predominantly stochastic emission from prompt convection behind the newly formed shock and from early convection within the PNS \citep{radice2019characterizing,vartanyan2023gravitational}. During the subsequent accretion phase, turbulent downflows striking the PNS excite its gravity and fundamental oscillation modes. The associated g- and f-mode tracks generally rises from a few hundred hertz to frequencies above 1\,kHz as the PNS contracts \citep{Morozova2018,Torres2019,vartanyan2023gravitational}. The SASI can add a lower-frequency, quasi-periodic component, commonly beginning near 100-200\,Hz. Its frequency evolves with the advection and acoustic timescales in the post-shock region and generally increases as a stalled shock retreats \citep{Blondin2003Sasi,Foglizzo2024,Hanke2013,Kuroda2016}. SASI activity occurs during the stalled-shock accretion phase and is suppressed once shock revival terminates accretion through the gain region. It may persist for longer in non-reviving or failed models than in successful explosions.

The rotational state of the progenitor core heavily influences the GW production. Slowly rotating or non-rotating cores produce mostly stochastic GW signals arising from convection, turbulence, and SASI activity, making waveform prediction challenging and detection via matched filtering difficult. Rapid rotation can instead introduce a pronounced prompt core-bounce signal through centrifugal deformation and can substantially enhance the GW amplitudes. However, the subsequent waveform morphology depends on the rotation profile and post-bounce dynamics. While non-axisymmetric rotational instabilities such as bar modes or low-T/|W| instabilities can develop \citep{kotake2017gravitational, abdikamalov2022gravitational}, many 3D rapidly rotating simulations show that the signal can retain a modal structure similar to non-rotating models (dominated by f- and g-modes), but with substantially larger GW amplitudes \citep{Dimmelmeier2008, powell2020,Andresen2021}.

To capture the broad phenomenology of GWs from CCSNe, we employed two complementary waveform catalogs derived using \texttt{Fornax}, a state-of-the-art radiation-hydrodynamics simulation framework. \texttt{Fornax} includes detailed microphysics, neutrino transport, and general relativistic effects to model CCSN dynamics. The two waveform sets are from \citet{radice2019characterizing} and \citet{vartanyan2023gravitational}, and are summarized in Table~\ref{tab:sne_horizons}. Together, they span a progenitor mass range of 9-25\,M$_\odot$ and provide complementary insights into different stages of GW emission. The simulations from \citet{radice2019characterizing} (Figure~\ref{fig:strainrad}) focus on the early to intermediate post-bounce evolution, spanning from 500 ms to just over one second. These simulations emphasize how the initial phases of GW emission are dominated by the g- and f- mode in the periphery of the PNS. The SASI feature is only observed in the more massive 25M$_{\odot}$ progenitor within this set.  

In contrast, the second set from \citet{vartanyan2023gravitational} extends the coverage into the late-time evolution, following the CCSN event beyond the prompt emission phase through to explosion or black hole formation. This longer timescale approach (up to several seconds post-bounce) captures continued PNS oscillations and late-time dynamical features that shape the gravitational waveform during the full duration of the event (Figure~\ref{fig:strainvar}).
These simulations indicate that the majority of GW energy is carried by the f- and g-mode oscillations of the PNS, whose characteristic frequency increases steadily as the PNS contracts over time. These modes are predominantly excited by accretion plumes striking the surface of the PNS, which not only stimulate coherent oscillations but also contribute to a broadband high-frequency component, correlated with the episodic nature of violent accretion. 
It is important to note that, within the simulations, the 12.25M$_\odot$ and 14M$_\odot$ models \citep{radice2019characterizing, vartanyan2023gravitational} do not undergo shock revival. Instead, the process of ongoing accretion ultimately results in the collapse of the object and the formation of a black hole. In contrast to successful explosions, BH-formers do not generate a high-frequency emission, rather, up to 95$\%$ of their total GW energy is concentrated directly in the f-mode. As the shock radius contracts below a critical value of approximately 100 km, a pronounced spiral-SASI emerges at a frequency of around 100–200 Hz. This is followed by a sudden stop of GW emission within less than a millisecond after the collapse of the black hole, resulting in a lack of residual low-frequency matter memory.

Together, these waveform catalogs cover a broad range of progenitor masses (9–25~M$_{\odot}$) and evolutionary stages, enabling a comprehensive analysis of GW emission across both prompt and late-time phases of CCSNe.

\section{GWFish and high-frequency sensitivity}
\label{sec:gwfish}

The core of our GW analysis is \texttt{GWFish}~\citep{dupletsa2023gwfish}\footnote{\url{https://github.com/janosch314/GWFish}}, a Fisher matrix-based simulation software designed for evaluating parameter estimation and detection capabilities for GW detectors. This code offers flexibility in incorporating various detector networks, noise power spectral densities (PSDs), and waveform models. Traditionally, \texttt{GWFish} was developed to study compact binary mergers, where the Fisher matrix formalism provides a reliable approximation for estimating parameter uncertainties under Gaussian noise assumptions.

However, GW signals from CCSNe present unique challenges due to their stochastic and complex nature, arising from turbulent hydrodynamics, neutrino-driven convection, and multi-dimensional instabilities (see \citet{Andresen:2017, vartanyan2023gravitational}). This complexity can invalidate standard Fisher matrix analyses, making direct parameter estimation via this method unreliable. Despite these limitations, \texttt{GWFish} includes a dedicated module for horizon estimation that can still yield valuable insights for CCSNe even with stochastic waveforms. In this context, the horizon distance is defined as the maximum luminosity distance at which a GW signal can be detected with a given signal-to-noise ratio (SNR) threshold. We define SNR as 
\begin{equation}\label{eq:snr_net}
    {\rm SNR}^2= 4 \int {\rm d} f \dfrac{|\tilde{h}(f)|^2}{S_{n}(f)}\,
\end{equation}
where $S_{n}(f)$ denotes the one-sided noise psd of the considered detector and $\tilde{h}(f)$ is the frequency domain signal projected onto the detector. We assume a minimum cut-off frequency of 1\,Hz.

Typically the SNR quantifies how well the observed data $d$ matches a given GW template $h$. In the ideal case of Gaussian noise without glitches, the SNR follows a $\chi^2$ distribution. However, real detector noise is non-Gaussian and contains glitches, complicating direct interpretation especially for low SNR. When forecasting detection horizons the precise noise realization and future glitches are unknown. Therefore, a typical SNR detection threshold is assumed to lie between 8 and 12, we take 10 as our threshold. 

A crucial aspect for signals in the kHz frequency range, which is characteristic of CCSN GWs, is the proper accounting for the detector transfer function. The long-wavelength approximation (LWA), commonly used in GW signal analysis, assumes a unity transfer function at all frequencies. Due to the limited speed of light, this approximation breaks down at characteristic frequencies $f_c = c / (2\pi L)$, where $L$ is the detector arm length. As detailed in \cite{Dhani:2025}, the detector transfer function deviates from unity at kHz frequencies, degrading the sensitivity of detectors at these frequencies. This is more relevant for proposed next-generation observatories like the ET compared to current-generation detectors. \texttt{GWFish} explicitly incorporates these high-frequency corrections, which are essential for accurate horizon estimation of high-frequency sources like CCSNe. 

Additionally, to enable horizon estimation for CCSNe, we have extended \texttt{GWFish}'s capabilities. This extension leverages our new time-series module to input time-domain CCSN waveforms from the hydrodynamical simulations introduced in Section~\ref{sec:gw} and to utilize the massive star spatial distribution described in Section~\ref{sec:progenitors} as the source population. 

For each star in the massive star distribution, we identify the closest progenitor mass from our catalogue of CCSN waveforms to approximate its GW signature, with the distance to each potential source derived from its position within the spatial distribution. This procedure enables us to combine realistic CCSN waveforms with a synthetic population of potential Galactic progenitors despite the currently limited number of publicly available simulations.
The expected SNR is then computed for each source and detector configuration. Since the detector response depends on the relative orientation between the source and the detector, which continuously change due to the Earth's rotation and orbital motion, we estimated the signal at random geocentric times. We can repeat this over the entire progenitor population spanning the MW, SMC and LMC to build to construct SNR distributions and derive the probability of GW detections for different detector networks. 

The SNRs computed here correspond to optimal waveform SNRs and should therefore be interpreted as sensitivity benchmarks rather than as the detection efficiencies of a realistic unmodeled CCSN search. External neutrino information can nevertheless constrain the relevant time-frequency region of the GW signal. For a SASI-dominated model, \citet{drago2023multimessenger} found that a neutrino-informed matched-filter search can increase the GW detection efficiency by up to approximately 30\% for sources within a few kpc assuming an Ligo-Virgo-Kagra \citep{aasi2015advanced,acernese2014advanced,akutsu2021overview} network. More general strategies combining low-energy-neutrino and GW searches have also demonstrated improved detection potential \citep{Halim2021}.

\section{Results}
\label{sec:results}

Using the framework described above, we applied {  \tt GWFish} to the waveforms introduced in Section~\ref{sec:gw} to estimate CCSN detection horizons. The MW detectability was quantified through SNR distributions obtained from simulated sources injected at sky locations drawn from the massive star distribution described in Section~\ref{sec:progenitors}. We also explored the performance of various detector network configurations, including the 10 km triangular ($\Delta$) and 15 km dual-L (2L) of the ET, both operating independently and in networks combined with a CE component. The US based CE is a proposed third-generation ground-based interferometric GW observatory, which design envisions two L-shaped facilities, one with 40 km arms and the other with 20 km arms, each hosting a single-detector configuration \citep{hall2022cosmic}. CE is optimized for high-frequency sensitivity and long-baseline precision, making it particularly effective for joint detections of CCSNe when operating in a network with ET. For the present analysis, we used only the CE 40km arms (CE40).
In Table \ref{tab:sne_horizons} we present the horizon estimates at a reference value of SNR = 10 for all waveform models and for four detector configurations, namely ET-$\Delta$, ET-2L, ET-$\Delta$+CE40, and ET-2L+CE40. While the horizon generally increases with progenitor mass, the trend is not strictly monotonic, as differences in the explosion dynamics and waveform morphology introduce model-dependent variations.

Across the MW, the spatial distributions of the simulated progenitors and potentially detectable events are therefore nearly coincident. The detection fraction is close to unity throughout the considered progenitor-mass range, so no significant mass-dependent selection bias is expected for Galactic CCSNe within the assumptions of our waveform sample.
Taking into account that the stellar disk of the MW extends to radii of approximately 15-20\,kpc, while the Large and Small Magellanic Clouds are located at distances of about 50 and 60 kpc respectively, the horizons reported in Table~\ref{tab:sne_horizons} imply a complete coverage of the Galactic CCSN population, but only partial coverage of the Magellanic Clouds for ET operating alone. 
This conclusion is broadly consistent with \citet{Choi2024}, who analyzed a suite of long-duration, initially non-rotating 3D \texttt{Fornax} models and likewise found that ET should detect the matter-motion GW signal from CCSNe throughout the Milky Way. Their analysis averages over the intrinsic viewing angle of the emission but assumes an optimally located source relative to the detector and adopts an SNR threshold of 8, whereas our calculation uses a \texttt{TRILEGAL}-based Galactic population, detector responses, an SNR threshold of 10, and explicit ET-CE network configurations.

\begin{figure}
    \centering
    \includegraphics[width=\linewidth]{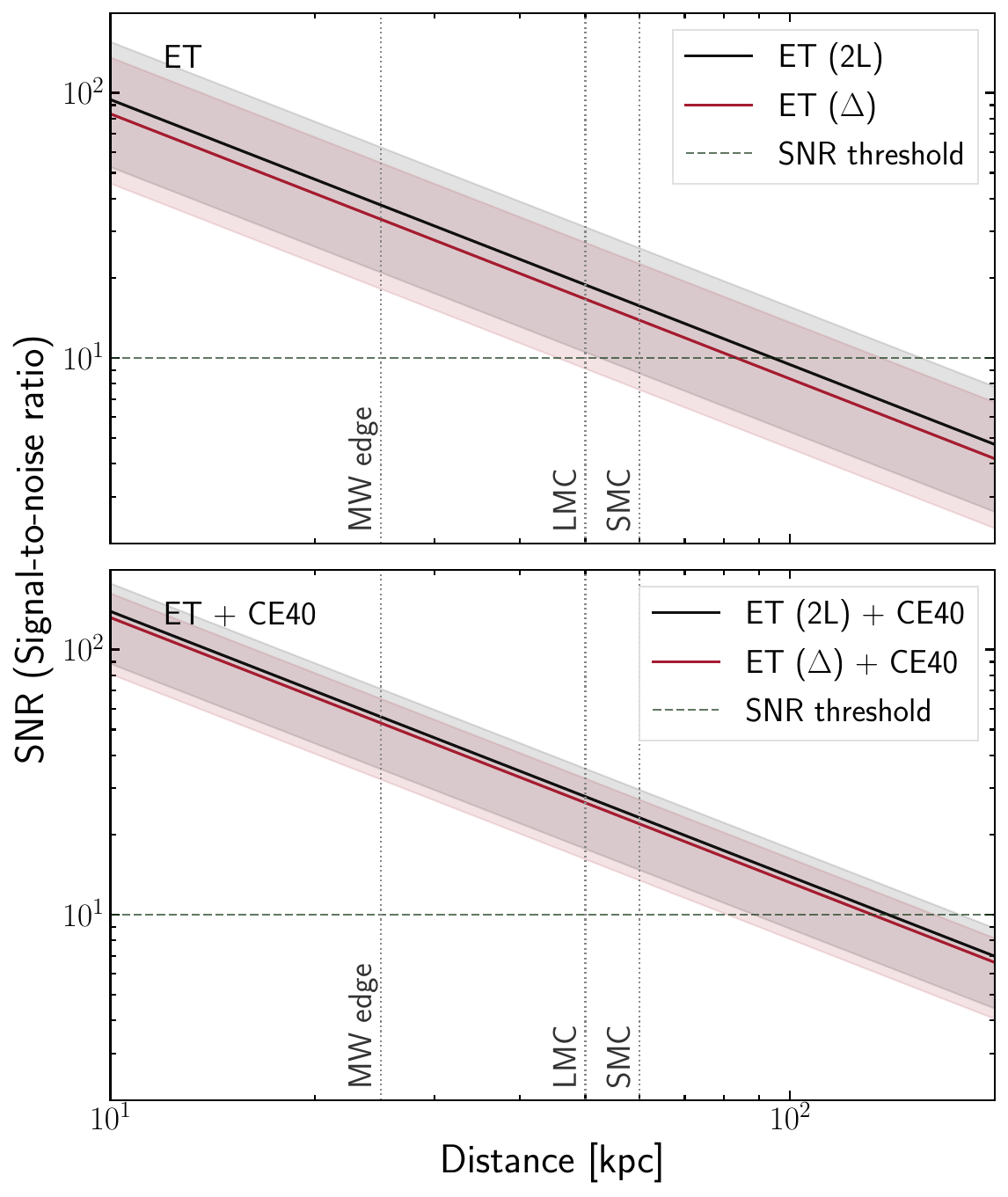}
    \caption{Signal-to-Noise Ratio as a function of distance for the 15 M$_{\odot}$ progenitor from \citep{radice2019characterizing}. In the top panel we compare the performances of the two ET designs, in the bottom panel the comparison is obtained adding in CE in the network. To guide the eye the Milky Way edge and the large and small Magellanic clouds are indicated.}
    \label{fig:snrvsdistance}
\end{figure}

For the most optimistic waveform models, the ET+CE40 network extends the accessible volume well beyond the Magellanic system, opening the possibility of multimessenger observations of CCSNe occurring in nearby satellite galaxies. Another insight that emerges from our results concerns the ET configuration. In Figure \ref{fig:snrvsdistance}, we compare the performance of the $\Delta$ and 2L layouts for the 15\,M$_{\odot}$ progenitor. The 2L configuration provides a sligh improvement of about 15-20$\%$ compared to the $\Delta$ when ET operates alone, and about 5–10$\%$ within a network including CE40. While this gain becomes less significant in the network case, where CE detectors compensate for marginal differences, it can still be impactful in standalone scenarios, especially for lower-mass progenitors near the threshold of detectability.

\begin{table*}[t!]
\centering
\caption{Detection horizons and properties of the considered detector networks and CCSN waveform sample.}
\begin{tabular}{l c c c c c}
\toprule
\multicolumn{1}{c}{Mass} & Duration & \multicolumn{2}{c}{ET [kpc]} & \multicolumn{2}{c}{ET+CE40 [kpc]} \\
\cmidrule(lr){3-4} \cmidrule(lr){5-6}
M$_\odot$ & (s) & 2L & Triangle & 2L & Triangle \\\midrule
{$9$}$^{b}$ & 1.04 & {$19.1^{+{12.4}}_{-{8.0}}$} & {$16.2^{+{10.2}}_{-{7.1}}$} & {$28.4^{+{7.2}}_{-{10.3}}$} & {$26.5^{+{5.0}}_{-{10.2}}$}
\\
\rowcolor{MidnightBlueLight}
{$9.25$}$^{a}$ & 2.75 & {$31.0^{+{19.6}}_{-{12.0}}$} & {$26.8^{+{16.8}}_{-{10.6}}$} & {$46.6^{+{10.5}}_{-{16.7}}$} & {$43.6^{+{8.0}}_{-{15.9}}$}
\\
{$9.50$}$^{a}$ & 2.14 & {$42.6^{+{28.0}}_{-{20.5}}$} & {$36.0^{+{23.0}}_{-{17.7}}$} & {$62.0^{+{19.9}}_{-{23.2}}$} & {$57.2^{+{19.3}}_{-{22.5}}$}
\\
\rowcolor{MidnightBlueLight}
{$10$}$^{b}$ & 0.76 & {$60.4^{+{38.6}}_{-{23.3}}$} & {$53.5^{+{33.8}}_{-{20.9}}$} & {$91.4^{+{20.6}}_{-{32.6}}$} & {$86.4^{+{16.3}}_{-{31.4}}$}
\\
{$11$}$^{a}$ & 3.08 & {$85.1^{+{55.7}}_{-{38.1}}$} & {$73.0^{+{46.2}}_{-{33.1}}$} & {$125.7^{+{36.4}}_{-{46.2}}$} & {$118.1^{+{32.3}}_{-{45.4}}$}
\\
\rowcolor{MidnightBlueLight}
{$12.25$}$^{a}$ & 2.01 & {$46.9^{+{30.1}}_{-{19.6}}$} & {$41.8^{+{26.3}}_{-{18.1}}$} & {$68.8^{+{17.8}}_{-{24.9}}$} & {$65.7^{+{12.9}}_{-{24.5}}$}
\\
{$13$}$^{b}$ & 0.50 & {$55.7^{+{36.0}}_{-{23.6}}$} & {$47.1^{+{29.9}}_{-{20.1}}$} & {$83.0^{+{21.5}}_{-{30.0}}$} & {$77.5^{+{16.4}}_{-{29.5}}$}
\\
\rowcolor{MidnightBlueLight}
{$14$}$^{a}$ & 2.49 & {$64.5^{+{42.9}}_{-{34.2}}$} & {$54.0^{+{35.0}}_{-{29.4}}$} & {$93.1^{+{36.3}}_{-{35.1}}$} & {$84.5^{+{36.2}}_{-{33.5}}$}
\\
{$15.01$}$^{a}$ & 3.80 & {$105.8^{+{67.5}}_{-{43.2}}$} & {$92.9^{+{58.6}}_{-{38.1}}$} & {$158.8^{+{38.8}}_{-{57.3}}$} & {$150.4^{+{29.0}}_{-{55.4}}$}
\\
\rowcolor{MidnightBlueLight}
{$19$}$^{b}$ & 0.86 & {$115.0^{+{73.2}}_{-{45.5}}$} & {$102.6^{+{64.7}}_{-{41.3}}$} & {$172.5^{+{40.5}}_{-{61.8}}$} & {$164.7^{+{30.0}}_{-{60.3}}$}
\\
{$23$}$^{a}$ & 4.20 & {$98.9^{+{64.6}}_{-{46.7}}$} & {$85.2^{+{54.5}}_{-{41.3}}$} & {$145.5^{+{44.2}}_{-{54.9}}$} & {$136.0^{+{41.3}}_{-{53.5}}$}
\\
\rowcolor{MidnightBlueLight}
{$25$}$^{b}$ & 0.61 & {$94.5^{+{61.6}}_{-{41.9}}$} & {$83.5^{+{52.8}}_{-{37.9}}$} & {$139.6^{+{38.4}}_{-{51.0}}$} & {$132.3^{+{31.0}}_{-{51.3}}$}
\\ \bottomrule
\end{tabular}
\tablefoot{Horizons are median detection distances for a network SNR threshold of 10 alongside the 95th and 5th percentiles, respectively, over random right ascension, declination, and geocentric times.\\
\tablefoottext{a}{Waveforms from \citet{vartanyan2023gravitational}.} 
\tablefoottext{b}{Waveforms from \citet{radice2019characterizing}.}}
\label{tab:sne_horizons}
\end{table*}

\begin{figure}[h!]
\centering
\includegraphics[width=0.5\textwidth]{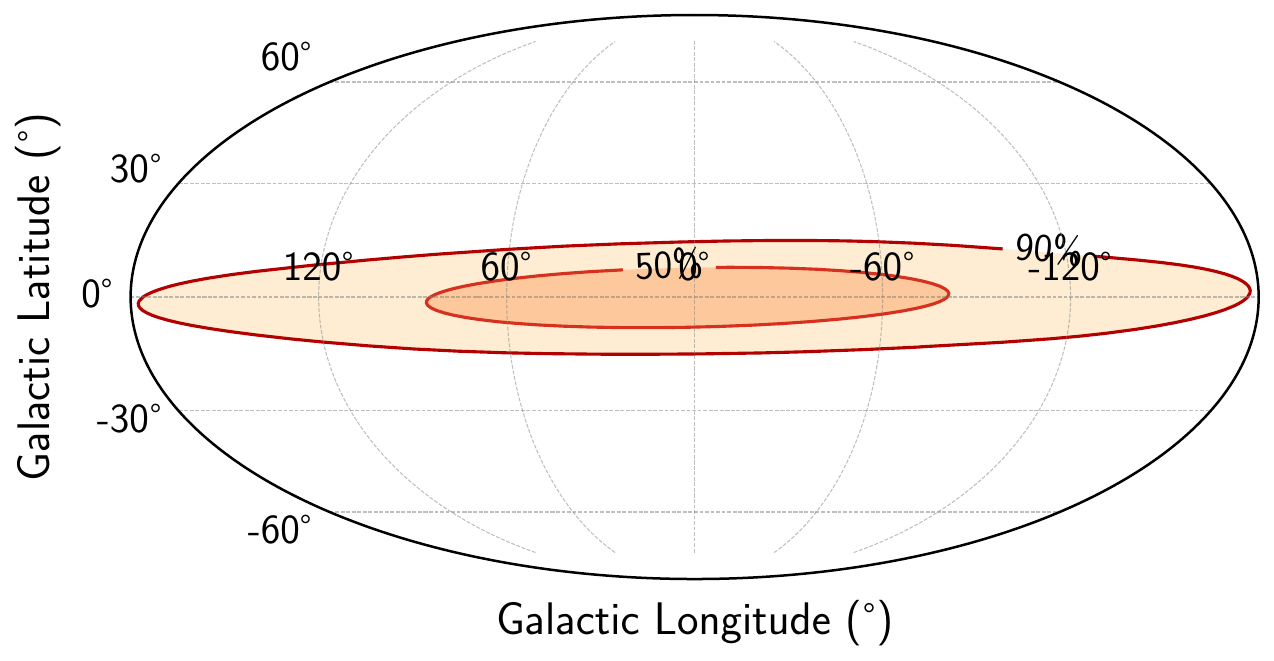}
\caption{Sky map showing the region where 50\% and 90\% of potential CCSN detections are. Assuming a ET (2L) configuration and $\text{SNR}\geq10$.}
\label{fig:snrmap}
\end{figure}

Figure \ref{fig:snrmap} maps on the Galactic plane the regions enclosing the 50\% and 90\% of the potentially detectable CCSN events, assuming a  ET (2L) configuration and a threshold of $\text{SNR} \ge 10$. As expected, the main detection probability areas are highly concentrated along the Galactic equator, reflecting the underlying distribution of massive progenitor stars within the Galactic disk.  
Section \ref{sec:discussion} discusses the implications of this complete coverage, particularly for events occurring in highly obscured regions of the Galactic plane.

\section{Discussion}
\label{sec:discussion}

\begin{figure}
    \centering
    \includegraphics[width=1\linewidth]{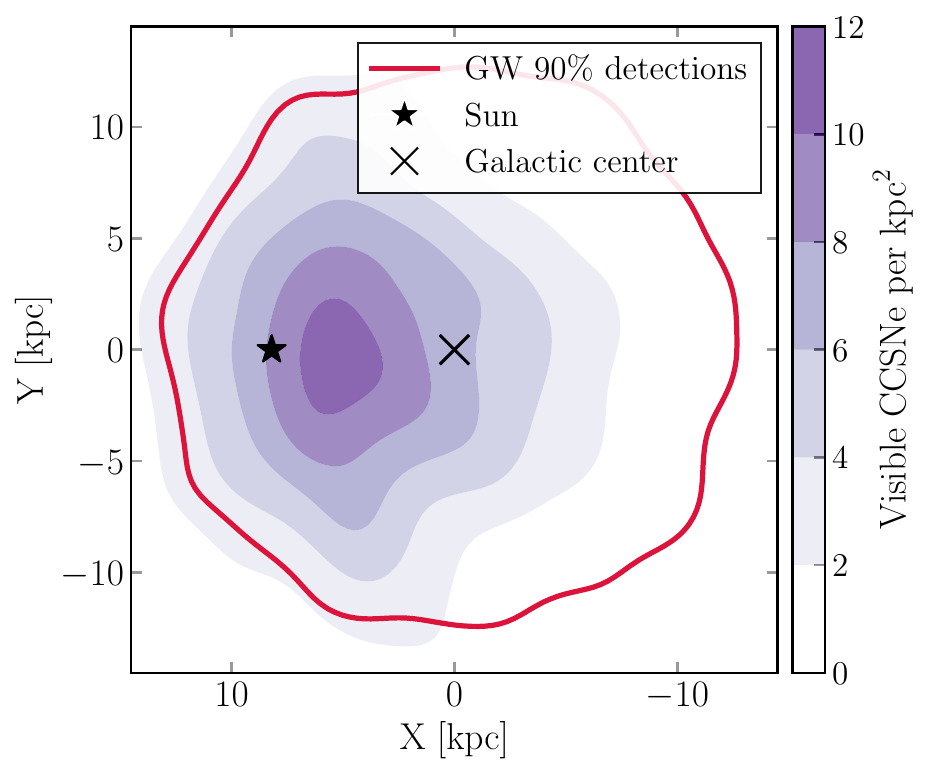}
    	\caption{Comparison between the GW reach and the spatial distribution of CCSNe visible from optical searches. The red contour shows $90\%$ of possible detections from an ET (2L) configuration, while the density plot highlights the regions with the highest probability of CCSN optical detection.}
	\label{fig:comparisonSNRvis}
\end{figure}

An important aspect emerging from our results is the role of GWs in overcoming observational biases introduced by dust extinction \citep{powell2025}. To investigate the detectability of CCSNe in the absence of electromagnetic counterparts,  we examined the Galactic dust extinction map from the Planck Collaboration \citep{abergel2014planck}. The dust distribution reveals that many GW-detectable regions are heavily obscured in optical wavelengths due to interstellar dust. To quantify this effect, we assigned a CCSN B-band absolute magnitude ($M_{B}$) to each simulated progenitor, randomly drawn from a Gaussian distribution with mean $M_{B}=-17$ mag and standard deviation $\sigma = 0.9$, based on the values from \cite{richardson2002comparative}. Extinction corrections for the apparent magnitudes in the MW \texttt{TRILEGAL} simulation are derived from the Planck thermal dust optical depth map at $353\text{ GHz}$ ($\tau_{353}$; \cite{abergel2014planck}). We converted this optical depth to the V-band extinction ($A_{V}$) adopting the relation $A_{V}=R_{V} (E(B-V)/\tau_{353}) \tau_{353}$ (where we assumed $E(B-V)/\tau_{353} = 1.49\times10^{4}$ mag) and finally scaled the extinction to the $B$-band using the Cardelli extinction law \citep{cardelli1989relationship}.
By applying the dust extinction map to the synthetic apparent magnitudes and assuming a limiting apparent magnitude of 22 (representative of the observing conditions in wide-field surveys like the Zwicky Transient Facility \citep{bellm2019}), we found that approximately $\sim 50\%$ of all potential Galactic CCSNe would be too heavily obscured to be detected optically, particularly toward the inner Galactic disk where dust density peaks (Figure~\ref{fig:comparisonSNRvis}).

To estimate the uncertainty associated with the intrinsic CCSN luminosity distribution, we repeated the same analysis assuming $M_B=-19$ mag and $M_B=-15$ mag, corresponding to the bright and faint ends of the CCSN population. This yields a visibility fraction of
\begin{equation}
f_{\rm vis} = 45\pm3\%
\end{equation}

\begin{figure}
    \centering
    \includegraphics[width=1\linewidth]{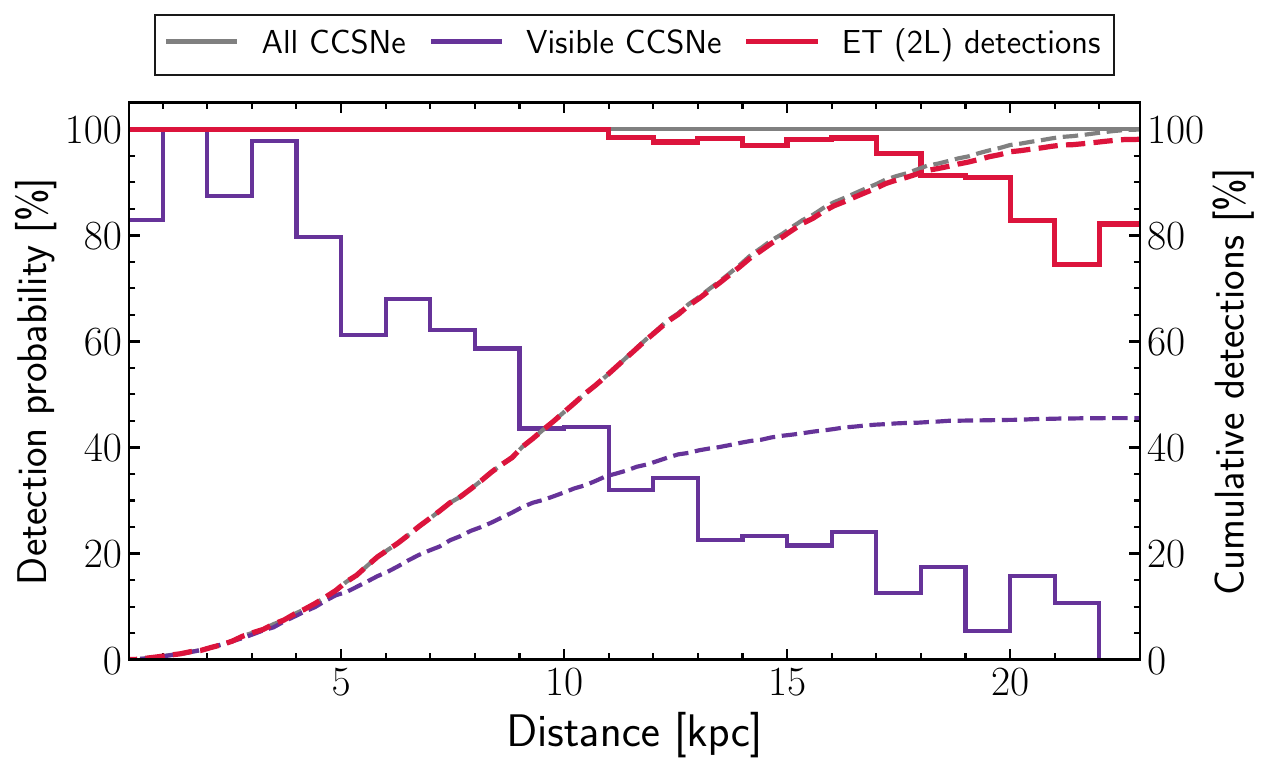}
    
    \caption{
    Probability density distributions (continuous lines) and cumulative distributions (dashed lines) of possible CCSNe (grey lines), CCSNe remaining optically visible given a limiting magnitude of 22 (purple lines), and gravitational-wave events detectable with a ET (2L) configuration at $\text{SNR} \geq 10$ (red lines).
    }
    \label{fig:distance_visibility}
\end{figure}

\begin{figure}
    \centering
    \includegraphics[width=\linewidth]{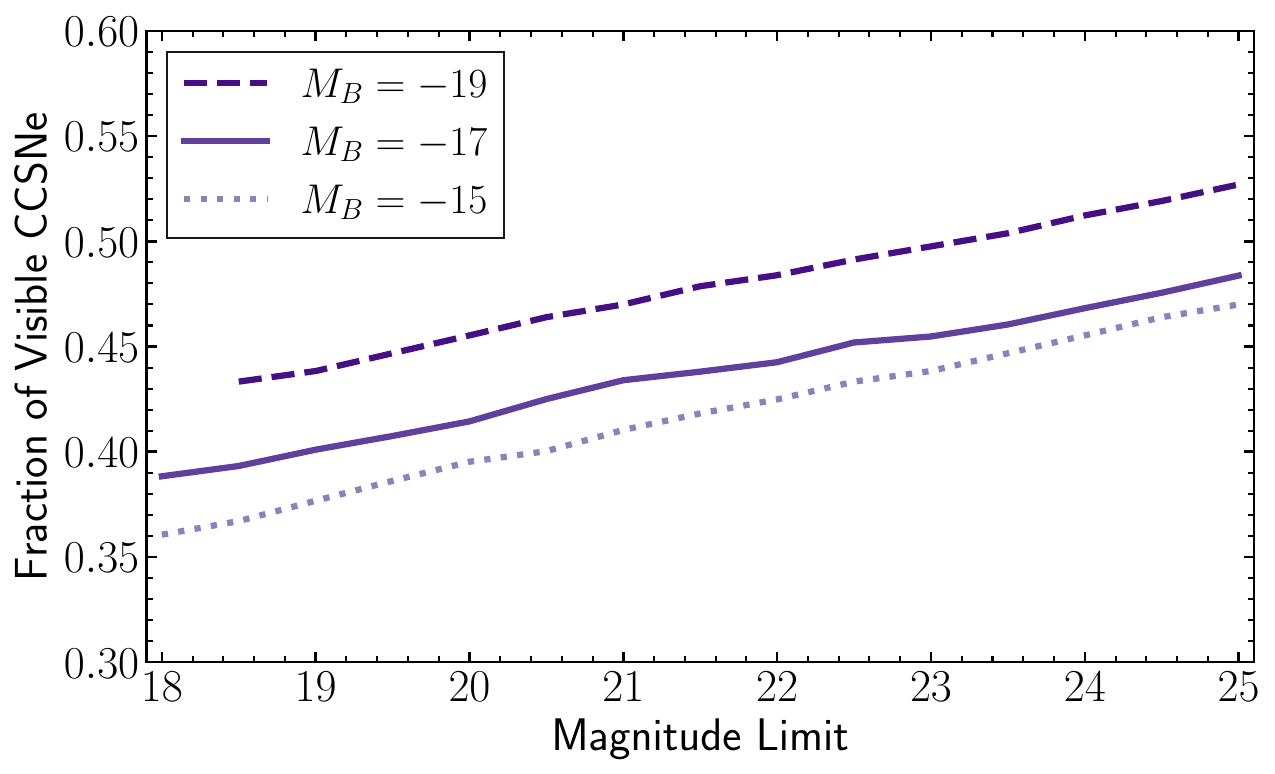}
    \caption{Fraction of visible CCSNe as a function of the survey magnitude limit. For different average intrinsic B-band absolute magnitudes.}
    \label{fig:magnitude_limit_visibility}
\end{figure}

This small variation indicates that the optical visibility is heavily dominated by dust extinction rather than the intrinsic CCSN luminosity, suggesting that a substantial fraction of Galactic CCSNe may remain optically hidden. In this context, GW observations provide a unique and unobstructed probe of the collapse process, independent of electromagnetic visibility, and become essential for identifying CCSNe that would otherwise be hidden in traditional optical surveys. The comparison between the GW reach and the subset of optically visible CCSNe shown in Figure~\ref{fig:comparisonSNRvis} further illustrates how ET retains high detection capability in regions where optical searches become incomplete. Figure~\ref{fig:distance_visibility} compares the cumulative fraction of CCSNe that remain optically visible with the fraction expected to be detectable through GWs by both ET configurations. While the fraction of optically observable events decreases rapidly with increasing distance because of interstellar extinction, the GW detection efficiency remains close to unity across most of the Galactic disk for signals with $\text{SNR}\geq10$, only declining toward the far side of the Galaxy. Finally, Figure~\ref{fig:magnitude_limit_visibility} shows how the visible fraction evolves as a function of the survey limiting magnitude. We see that even large differences in magnitude limit of intrinsic magnitudes does not significantly increase the fraction of visible CCSNe.


These findings further emphasize the importance of multimessenger strategies combining GWs, neutrinos and electromagnetic observations for obtaining a complete census of Galactic CCSNe.
Despite these promising results, some limitations must be acknowledged. The detection horizons presented in Table~\ref{tab:sne_horizons} are derived from a limited sample of publicly available 3D CCSN simulations \citep{vartanyan2023gravitational, radice2019characterizing}. We deliberately restrict our waveform sample to simulations performed with the \texttt{FORNAX} framework in order to maintain a homogeneous treatment of hydrodynamics, neutrino transport, microphysics and GW extraction across the progenitor-mass range considered. However, although the \texttt{FORNAX} code incorporates state-of-the-art neutrino-hydrodynamics, independent 3D simulation codes exhibit notable differences in post-bounce accretion dynamics, shock recovery times, and PNS convection \citep{powell2019gw, Andresen2021}. Consequently, alternative simulation codes predict slightly higher GW amplitudes or shifts in the $g$- and $f$-mode frequency trajectories, which would systematically alter the predicted ET detection horizons and SNR distributions.
Therefore, our findings should be interpreted as a demonstration of capability under a plausible scenario rather than an exhaustive prediction. The continued development of more sophisticated simulations is essential for building the comprehensive template banks needed for robust signal detection and parameter estimation.
Furthermore, our primary waveform sample is restricted to non-rotating progenitors. Rapidly rotating or magnetorotational core collapses can drive significantly stronger, coherent GW signals. 3D magnetohydrodynamic simulations of rotating and magnetized cores \citep{ Obergaulinger2021, schnauck2025} show that rotation and magnetic fields may modulate the explosion dynamics, inducing deterministic prompt core-bounce bursts followed by enhanced non-axisymmetric emissions. In particular, magnetorotational models for high-mass progenitors ($25-35\,\mathrm{M}_\odot$) indicate that third-generation observatories like the ET could detect such energetic events up to extragalactic distances of $\sim10\,\mathrm{Mpc}$ \citep{schnauck2025}, substantially expanding the observational horizon compared to non-rotating scenarios.
In addition, deterministic waveform components from rotation would facilitate Fisher-matrix-based parameter estimation within \texttt{GWFish}, allowing forecasts for constraints on progenitor angular momentum, core structure, and the nuclear equation of state.	Similarly, our sample spans progenitor masses only up to $25\,\mathrm{M}_\odot$. However, more massive progenitors ($\gtrsim 25-40\,\mathrm{M}_\odot$), which frequently fail to explode and undergo direct collapse to a black hole, produce violent post-bounce accretion and prolonged SASI activity \citep{ powell2025, uchida2019gravitational}. Recent 3D studies demonstrate that such high-mass or failed events generate enhanced high-frequency GW emission that could extend the ET horizon out to several megaparsecs. Anisotropic neutrino emission can additionally produce a long-duration GW memory effect through the direction-dependent flux of radiated energy \citep{vartanyan2023gravitational,Choi2024}, with a similar but weaker contribution predicted from asymmetric matter ejection. In long-duration 3D simulations, the strain amplitude of this neutrino-memory component can exceed that of the oscillatory matter signal at low frequencies ($\lesssim 10-30\rm  Hz$). While this would substantially boost SNR which is especially relevant for extragalactic detection prospects, our waveforms already achieve near-complete detection coverage across the simulated Galactic progenitor population (see Section \ref{sec:results}). Consequently, the inclusion of neutrino memory would likely improve detection confidence for MW events without altering the core findings of this study. 

Despite these uncertainties in the GW signal alone, the true potential of ET will be realized through multimessenger astronomy. This is where the synergy with neutrino observatories can be relevant. The neutrino burst associated with a CCSN is expected to precede the electromagnetic signal and arrive nearly simultaneously with the GW signal. In the case of partially obscured CCSNe, joint observations by GW and neutrino observatories can provide sky localizations with accuracies of a few square degrees. This precision allows wide-field instruments such as the Vera C. Rubin Observatory \citep{2019ApJ...873..111I} to efficiently search for and identify the optical counterpart. Conversely, in high-extinction scenarios, the rapid sky localization provided by this joint network becomes essential to guide deep, pointed follow-up observations by infrared or high-energy observatories, which can  identify highly obscured events.

For neutrinos, a global network such as the Supernova Early Warning System (SNEWS) will provide high-confidence, low-latency alerts \citep{Al_Kharusi_2021} by combining detections from geographically separated facilities, capturing the massive low-energy MeV neutrino burst that carries away the bulk of the gravitational binding energy  \citep{Scholberg:2012}. Individual detectors such as Hyper-Kamiokande \citep{abe2018hyper}, DUNE \citep{DUNE:2019}, IceCube \citep{Griswold:2025}, JUNO \citep{ElHedri:2025}, KM3NeT, and LNGS dark-matter experiments \citep{Pagliaroli:2024} will offer complementary flavor, timing, and pointing capabilities, which together enable triangulation and early warnings from neutrino data \citep{Kate:2019,Colomber:2020}.

\section{Conclusion}\label{sec:conclusion}
We assessed the capability of the ET to detect CCSN GWs using a self-consistent framework that combines \texttt{TRILEGAL} distributions with state-of-the-art 3D PNS-driven CCSN waveform models. Our results show that ET will provide essentially complete coverage of CCSNe occurring within the MW, while its coverage of the Magellanic Clouds depends strongly on the detector configuration and progenitor waveform.
Depending on the progenitor properties and waveform morphology, ET alone reaches detection horizons between $\sim$20 and $\sim$115\ kpc (90$\%$ confidence level), while a network including ET and CE can extend the reach up to $\sim$170 kpc. These detection horizons are comparable with previous estimates based on {\tt CoCoNuT-FMT} simulations. \cite{powell2019gw} found ET detection distances of $\sim180\,\rm kpc$ for their 18 M$_{\odot}$ progenitor model, using a Bayesian analysis. Our predicted range of $\sim20-115\,\rm kpc$ is of the same order, with the differences reflecting the adopted waveform set and detection criteria. The range found here should therefore be interpreted as representative of the non-rotating, PNS-driven waveform population adopted in this work, rather than as a single characteristic or upper-limit distance for ET CCSN detections.

For the representative 15\,M$_{\odot}$ progenitor considered, the ET 2L configuration achieves a horizon distance of $\sim$106\,kpc, ensuring a high-probability detection of Galactic CCSNe. Regarding the detection prospects up to the Magellanic-Cloud, assuming a progenitor population with a mass distribution similar to that adopted for the Milky Way, the fraction detectable out to the SMC is approximately 20\% with ET-2L and 6\% with ET-$\Delta$, increasing to more than 70\% for either configuration when CE40 is included.

We also find that the 15 km dual-L configuration provides a performance improvement of approximately 15-20\,$\%$ over the triangular design in standalone operation, although this difference decreases to 5-10\,$\%$ when ET is part of a global detector network alongside CE.
Using the same SFR adopted for progenitor simulation and exploiting its relation with the CCSNR, we found a Galactic rate of $\sim$1.9 (100 yr)$^{-1}$ events. Combined with the detection horizons achieved by ET, this result implies that the detector is expected to observe nearly all possible Galactic core-collapse events occurring during its operational lifetime. Assuming our inferred Galactic CCSN rate of $1.9\,(100\,{\rm yr})^{-1}$, a detection efficiency close to unity, and 20 years of effective ET operation, Poisson statistics imply a probability of approximately $32\%$ of detecting at least one Galactic CCSN. Importantly, our analysis also shows that a large fraction of these events may remain optically hidden. By modeling Galactic dust extinction, we estimate that only $45\% \pm 3\%$ of CCSNe would be detectable by optical surveys with a limiting magnitude of 22, implying that more than half of the Galactic SN events, particularly those occurring toward the heavily obscured inner disk, could be hidden in optical searches. In this context GWs, together with neutrinos, become fundamental to identify core-collapse events, showing how these galactic events are a unique opportunity for multimessenger astronomy.
Overall, ET will represent a transformative step for CCSN multimessenger astronomy as it will provide a high detection efficiency for the next Galactic core-collapse event occurring during detector operation, regardless of dust extinction, and an unprecedented window into the physics of stellar collapse and the chemical evolution of our Galaxy.

\emph{Software}:
\texttt{matplotlib}\citep{hunter2007matplotlib}, \texttt{numpy}\citep{harris2020array}, \texttt{GWFISH} \citep{dupletsa2023gwfish}, \texttt{TRILEGAL}\citep{girardi2005star,girardi2012trilegal}

\begin{acknowledgements} 
IFG and MB acknowledge support from the Astrophysics Center for Multi-messenger Studies in Europe (ACME), funded under the European Union’s Horizon Europe Research and Innovation Program, Grant Agreement No. 101131928.  IFG acknowledges financial support from Piano Nazionale di Ripresa e Resilienza (PNRR) through the ETIC – Einstein Telescope Infrastructure Consortium (Missione 4/Componente 2/Investimento 3.1 – IR0000004).
\end{acknowledgements}

\bibliographystyle{aa}
\bibliography{reference} 

\end{document}